\pdfoutput=1
\documentclass[11pt,twoside,a4paper,cmspaper,final,collab]{cms-tdr}
\def\svnVersion{74cec2d}\def\svnDate{2023/11/20}\def\cmsCernNoTag{CERN-EP-2022-164}\def\cmsCernDate{\today}\def\cmsMessage{Published in Physical Review D as \href{http://dx.doi.org/10.1103/PhysRevD.108.092004}{\doi{10.1103/PhysRevD.108.092004}.}}
\begin{document}\cmsNoteHeader{HIN-18-019}

\RCS$Revision$
\RCS$HeadURL$
\RCS$Id$

\newcommand{\pPb}{\ensuremath{{\Pp}\text{Pb}}\xspace}
\newcommand{\Pbp}{\ensuremath{\text{Pb}{\Pp}}\xspace}

\newcommand{\Pom}{\ensuremath{\text{I\!P}}\xspace}
\newcommand{\PomPb}{\ensuremath{\Pom\text{Pb}}\xspace}
\newcommand{\PomP}{\ensuremath{\Pom\Pp}\xspace}
\newcommand{\PhotonP}{\ensuremath{\gamma\Pp}\xspace}
\newcommand{\PhotonPb}{\ensuremath{\gamma\text{Pb}}\xspace}
\newcommand{\PomPPhotonP}{\ensuremath{\PomP}+\ensuremath{\PhotonP}\xspace}

\newcommand{\dEtaF}{\ensuremath{\Delta\eta^{\text{F}}}\xspace}
\newcommand{\dSigmadetaF}{\ddinline{\sigma}{\dEtaF}\xspace}
\newcommand{\dSigmadetaFFrac}{\dd{\sigma}{\dEtaF}\xspace}

\newcommand{\Aepsilon}{\ensuremath{A\epsilon}\xspace}

\newcommand{\EPOS}{\textsc{epos-lhc}\xspace}
\newcommand{\QGSJET}{\textsc{qgsjet~ii}\xspace}

\newcommand{\ZDCm}{\ensuremath{\mathrm{ZDC-}}\xspace}
\newcommand{\HFm}{\ensuremath{\mathrm{HF}-}\xspace}
\newcommand{\HFp}{\ensuremath{\mathrm{HF}+}\xspace}

\hyphenation{ATLAS}
\hyphenation{HIJING}
\hyphenation{QGSJET}

\cmsNoteHeader{HIN-18-019}

\title{First measurement of the forward rapidity gap distribution in \texorpdfstring{\pPb collisions at
    $\sqrtsNN = 8.16\TeV$}{pPb collisions at sqrt(s[NN]) = 8.16 TeV}}

\date{\today}

\abstract{
  For the first time at LHC energies, the forward rapidity gap spectra from proton-lead collisions for both proton and lead dissociation processes are presented. The analysis is performed over 10.4 units of pseudorapidity at a center-of-mass energy per nucleon pair of $\sqrtsNN = 8.16\TeV$, almost 300 times higher than in previous measurements of diffractive production in proton-nucleus collisions.
  For lead dissociation processes, which correspond to the pomeron-lead event topology, the \EPOS generator predictions are a factor of two below the data, but the model gives a reasonable description of the rapidity gap spectrum shape. For the pomeron-proton topology, the \EPOS, \QGSJET, and \HIJING predictions are all at least a factor of five lower than the data. The latter effect might be explained by a significant contribution of ultra-peripheral photoproduction events mimicking the signature of diffractive processes.
  These data may be of significant help in understanding the high energy limit of quantum chromodynamics and for modeling cosmic ray air showers.
}

\hypersetup{
  pdfauthor={CMS Collaboration},
  pdftitle={First measurement of the forward rapidity gap distribution in pPb collisions at sqrt(s[NN]) = 8.16 TeV},
  pdfsubject={CMS},
  pdfkeywords={CMS, diffraction, proton-lead collisions}
}

\maketitle

\section{Introduction}

Diffractive scattering, \ie a process when there is an exchange of an object with vacuum quantum numbers between colliding particles~\cite{Gribov:1961ex,Chew:1961ev}, is one of the main topics of high energy hadronic physics.
The neutral colorless object that plays the main role in such exchanges, known as the \textit{pomeron} (\Pom)~\cite{Low:1975sv,Nussinov:1975mw,Fadin:1975cb}, is deeply connected to the fundamental nature of quantum chromodynamics (QCD).
In spite of many decades of study, diffractive processes are still poorly understood \cite{Kaidalov:1979jz, Donnachie:1984xq, N.Cartiglia:2015gve}.

The space-time development of hadronic diffractive processes inside a nucleus, because of intranuclear pomeron interactions, may involve rather specific interference effects due to Gribov inelastic screening contributions~\cite{Gribov:1968jf}.
Such contributions are absent in the standard Glauber approach, which only includes elastic intranuclear rescatterings~\cite{Glauber:1959}.
It has also been suggested that the nucleus can serve as a color filter for different transverse states of an incident hadron in diffractive hadron-nucleus scatterings~\cite{Kopeliovich:1981pz}.
Such interactions could manifest a new color dynamics~\cite{Kopeliovich:1981pz, x15} in the form of inelastic Gribov contributions, which may be detectable using the dependence of diffraction on the nuclear mass, A~\cite{x15}.

At center-of-mass energies of several tens of \GeV the HELIOS and EHS/NA22 Collaborations found that the cross section for hadron-nucleus diffraction scales as ${\simeq}A^{1/3}$~\cite{HELIOS,EHS}.
This suggests that, for these energies, hadron-nucleus diffraction occurs mostly in peripheral collisions.
In contrast, inelastic interactions are expected to follow black-disk scattering, with a cross section scaling as ${\simeq}A^{2/3}$.
Although there are a number of different models for soft proton-nucleus interactions within the general Glauber--Gribov approach~\cite{x14, x15, x16, x17, Deng:2010mv, Pierog:2013ria, Ostapchenko:2010vb}, and some of them are implemented in event generators~\cite{Deng:2010mv, Pierog:2013ria, Ostapchenko:2010vb}, there are no experimental data available for comparison at higher energies.
This paper presents the first data on proton-nucleus diffraction at the CERN LHC.

Diffractive processes at high energies are characterized by large gaps in the rapidity distribution of final-state particles, while the probability to find a continuous rapidity region $\Delta\eta$ free of particles is suppressed exponentially in nondiffractive inelastic events, $\ddinline{\sigma}{\Delta\eta} \sim \exp({-\Delta\eta})$.
Three major types of diffractive processes can be selected by the topology of the events~\cite{Barone:2002cv,Kaidalov:1979jz}.
Single-diffractive processes are characterized by an intact projectile hadron and a large forward rapidity gap (FRG).
This gap extends from the projectile rapidity back towards the region populated by products from the inelastic interaction between the emitted pomeron and target hadron.
In double-diffractive events, the hadron emitting a pomeron dissociates as well, so the rapidity gap appears between the two groups of dissociation products.
Central-diffractive processes, or double pomeron exchanges, are characterized by particles in the central rapidity region from a pomeron-pomeron interaction, surrounded by extended rapidity gaps.

Experimentally, a rapidity gap candidate event can be observed as an inelastic collision event with a continuous region devoid of detected final-state particles.
To increase the acceptance for single-diffractive dissociation, the search for pseudorapidity gaps, \dEtaF, starts from the most forward region of a detector.
Both the ATLAS~\cite{ATLASFRG} and CMS~\cite{Robert} Collaborations have measured the FRG cross section, $\ddinline{\sigma}{\dEtaF}$, for proton-proton (pp) collisions at a center-of-mass energy of $\sqrt{s} = 7\TeV$.

In high energy proton-lead (\pPb) collisions, events with large rapidity gaps may originate not only from the pomeron exchange process, but also from ultra-peripheral photoproduction~\cite{Baltz:2007kq}.
The electromagnetic field of a nucleus results in an equivalent flux of quasi-real photons which scales as $Z^2$.
The topology of photon-proton collisions is indistinguishable from those induced by pomeron exchange.

Diffraction of hadrons by nuclear targets at very high energies is also relevant for cosmic-ray physics~\cite{Luna:2004eg}.
Diffractive processes have a large cross section, and thus significantly contribute to the hadronic interactions that define the development of extensive air showers from high energy cosmic rays.
A correct realization of the mechanism of inelastic diffraction is at present one of the most difficult tasks for hadronic generators used in shower simulations~\cite{Ostapchenko:2014mna}.
Data on diffraction in \pPb collisions can provide valuable input for checking and tuning these generators.

We present in this paper, for the first time at LHC energies, FRG distributions in \pPb collisions for both pomeron-lead and pomeron-proton topologies.
The analysis is performed over 10.4 units of pseudorapidity at a center-of-mass energy per nucleon pair of $\sqrtsNN = 8.16\TeV$, \ie, almost 300 times higher than in the previous measurements of diffractive production in proton-nucleus collisions~\cite{HELIOS,EHS}.
The data are compared to the \HIJING~v2.1~\cite{Deng:2010mv}, \EPOS~\cite{Pierog:2013ria} and \QGSJET-04~\cite{Ostapchenko:2010vb} event generators.
The latter two generators are both based on the Gribov--Regge theory~\cite{Gribov:1967vfb,Barone:2002cv,Kaidalov:1979jz}, implementing it in different ways.
The \QGSJET generator is a theory driven model, in which multi-pomeron interactions are accounted for with resummation of relevant diagrams, while \EPOS relies more on data driven parameterizations.
The \HIJING generator is based on \PYTHIA~\cite{Sjostrand:1987su}, with semi-hard parton scatterings described by perturbative QCD, and with soft interactions modeled by string excitations with an effective cross section.
In particular, \HIJING has no low-mass diffractive excitations, which are implemented in the \EPOS and \QGSJET generators.

Tabulated results are provided in the HEPData record for this analysis~\cite{HEPData}.

\section{Experimental setup}
The central feature of the CMS apparatus is a superconducting solenoid of 6\unit{m} internal diameter, providing a magnetic field of 3.8\unit{T}.
Within the solenoid volume are a silicon pixel and strip tracker, a lead tungstate crystal electromagnetic calorimeter (ECAL), and a brass and scintillator hadron calorimeter (HCAL), each composed of a barrel and two endcap sections. Muons are measured in gas-ionization detectors embedded in the steel flux-return yoke outside the solenoid.  The silicon detectors provide tracking in the region $\abs{\eta} < 2.5$,  ECAL and HCAL cover the region  $\abs{\eta} < 3.0$, while the muon system covers the region  $\abs{\eta} < 2.4$.
In the forward region, the Hadron Forward (HF) calorimeters cover the region $2.85 < \abs{\eta} < 5.19$.
The HF calorimeters are made of steel absorber with longitudinal quartz fibers, so that bundles of quartz fibers form towers with individual readout, and provide a fine transverse segmentation with a typical tower size of $0.175{\times}0.175$ ($\Delta\eta{\times}\Delta\phi$, where $\phi$\ is azimuthal angle in radians).

Events of interest are selected using a two-tiered trigger system. The first level, composed of custom hardware processors, uses information from the calorimeters and muon detectors to select events at a rate of around 100\unit{kHz} within a fixed latency of about 4\mus~\cite{Sirunyan:2020zal}. The second level, known as the high-level trigger, consists of a farm of processors running a version of the full event reconstruction software optimized for fast processing, and reduces the event rate to around 1\unit{kHz} before data storage~\cite{Khachatryan:2016bia}.
A more detailed description of the CMS detector can be found in Ref.~\cite{Chatrchyan:2008zzk}.

Analysis in the midrapidity region $\abs{\eta} < 3.0$ is based upon objects produced by the
CMS particle-flow (PF) algorithm~\cite{Sirunyan:2017ulk}, which reconstructs and identifies each individual particle-flow candidate with an optimized combination of information from the various elements of the CMS detector.
The energy of electrons is determined from a combination of the electron momentum determined by the tracker, the energy of the corresponding ECAL cluster, and the energy sum of all bremsstrahlung photons spatially compatible with originating from the electron track. The energy of muons is obtained from the curvature of the corresponding track. The energy of charged hadrons is determined from a combination of their momentum measured in the tracker and the matching ECAL and HCAL energy deposits, corrected for the response function of the calorimeters to hadronic showers. Finally, the energy of neutral hadrons is obtained from the corresponding corrected ECAL and HCAL energies.

To compare the reconstructed data to \EPOS and \HIJING generator predictions, a detailed Monte Carlo (MC) simulation of the CMS detector response, based on the \GEANTfour framework~\cite{Agostinelli:2002hh}, is applied to the generated events. The simulated events are then processed and reconstructed in the same way as the collision data.

\section{Event selection}
The analysis is performed using proton-lead (\pPb) collisions, in which the proton moves counterclockwise and the lead ion moves clockwise, and lead-proton (\Pbp) collisions, in which the directions of the two beams are reversed.
In the CMS coordinate system, the direction of the proton beam in \pPb collisions defines positive rapidity.
Since the rigidity, \ie, the ratio of momentum to charge,  but not the energy of the proton and lead beams is the same, the nucleon-nucleon center-of-mass is found at a laboratory frame rapidity of $y_{\text{lab}}={\pm}0.465$, depending on the lead beam direction.
The data were recorded in the fall of 2016 and correspond to an integrated luminosity of 6.4\mubinv~\cite{CMS:2018fkg}.

For this analysis, events are selected with several hardware-based triggers.
The zero bias (ZB) trigger  only requires the presence of proton and lead bunches in the CMS detector.
The beam bunches are detected by induction counters placed 175\unit{m} from the interaction point on each side of the experiment.
The minimum bias (MB) trigger requires the presence of proton and lead beams
and an energy deposit of ${\geqslant}7\GeV$ in at least one of the HF calorimeters.
The analysis is performed with the minimum bias events while the zero bias data sets were used for detector acceptance corrections and in studies of systematic effects. In addition, a set of events triggered on non-colliding bunches is used to study the noise in the detector.

The integrated luminosities of the minimum bias data sets are 3.9 and 2.5\mubinv for \Pbp and \pPb, respectively, with a total uncertainty of 3.5\%~\cite{CMS:2018fkg}. The analysis is performed on each data subset separately, and the discrepancy between the obtained results is used as a measure of the systematic uncertainty due to the detector effects which are not accounted for in the MC simulation. The final distributions are obtained as a combination of the two results.

The offline selection requires that at least one tower in either HF has an energy deposit of at least 10\GeV.
The edge towers of the HF, $2.85 < \abs{\eta} < 3.14$, are shadowed by the endcap of the Hadronic Calorimeter and thus are not considered in the selection.
Excessively noisy towers, whose internal noise significantly exceeds the average noise level of the HF towers are also excluded.
Events containing more than one candidate vertex~\cite{CMS_TR} found close to the nominal interaction point are rejected from the analysis in order to minimize a contribution from simultaneous inelastic collisions.
Events are required to have a vertex that is within 15\cm of the nominal interaction point along the beam axis and with 1\cm perpendicular to the beam axis.

\begin{figure*}[htb]
  \centering
  \includegraphics[width=0.27\textwidth]{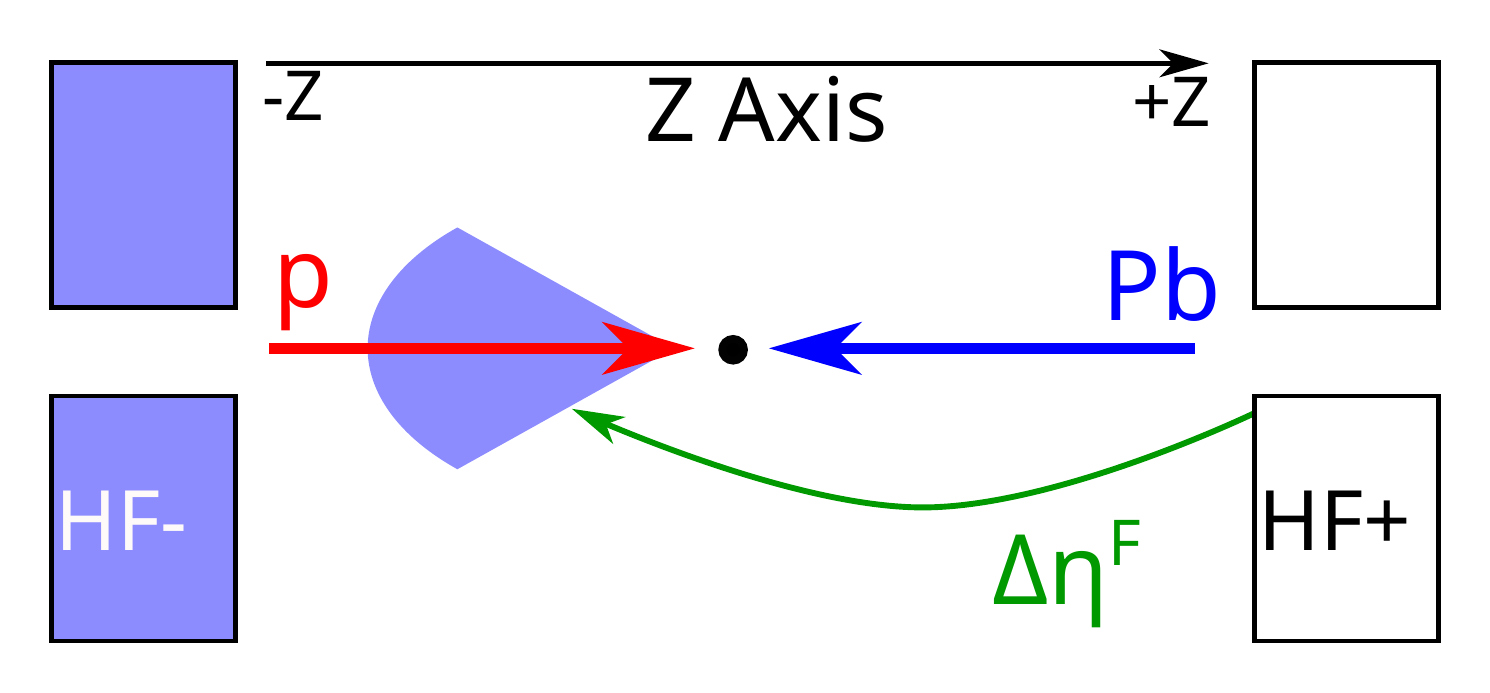}
  \includegraphics[width=0.21\textwidth]{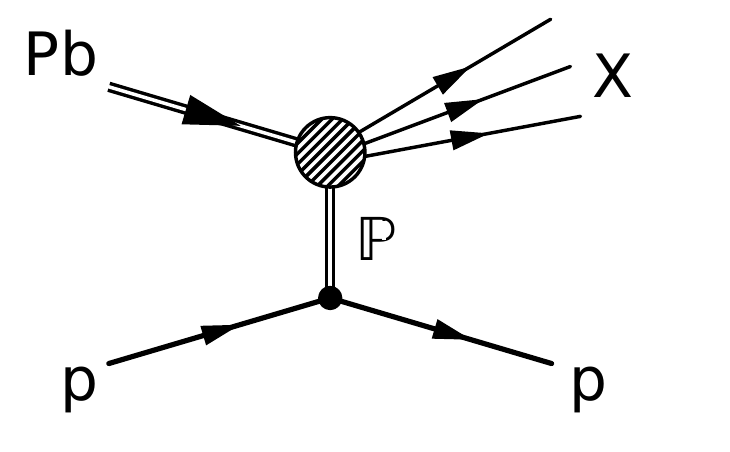}
  \vline \vspace*{5pt}
  \includegraphics[width=0.27\textwidth]{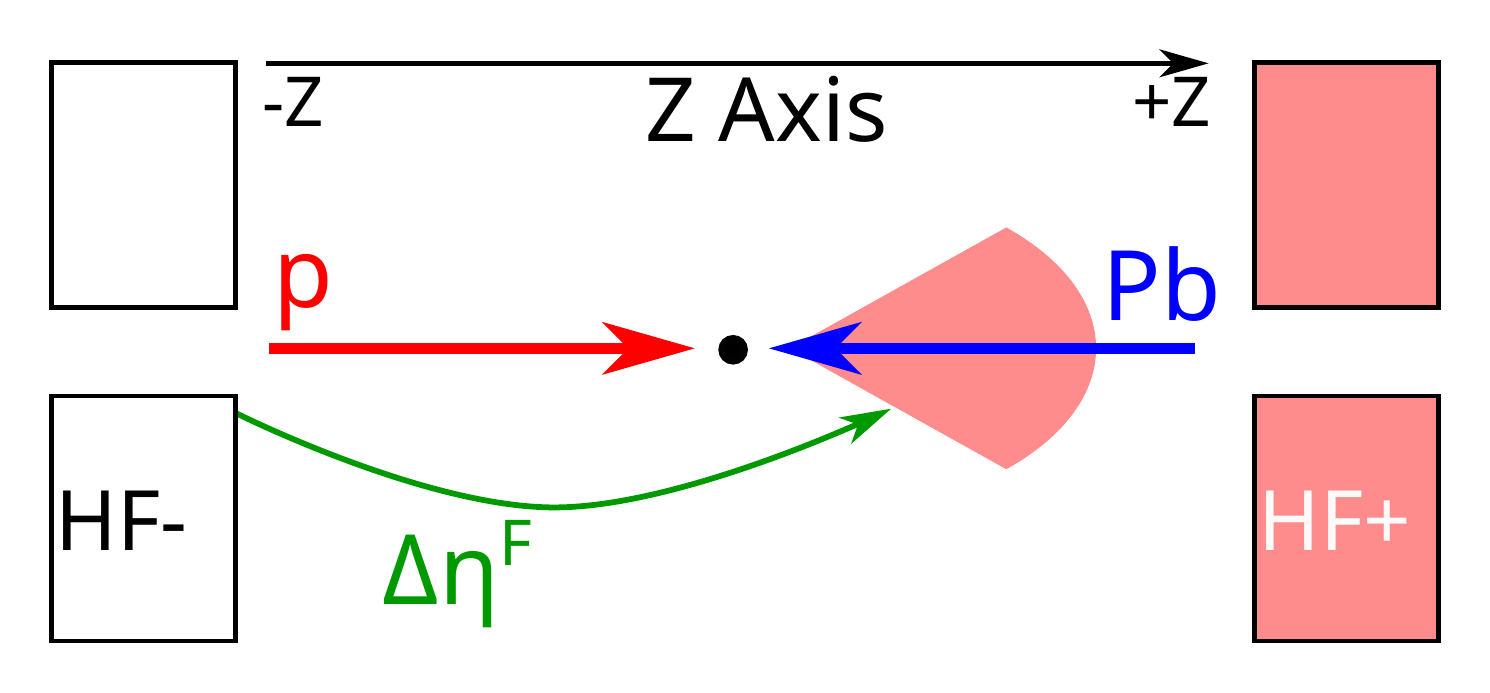}
  \includegraphics[width=0.21\textwidth]{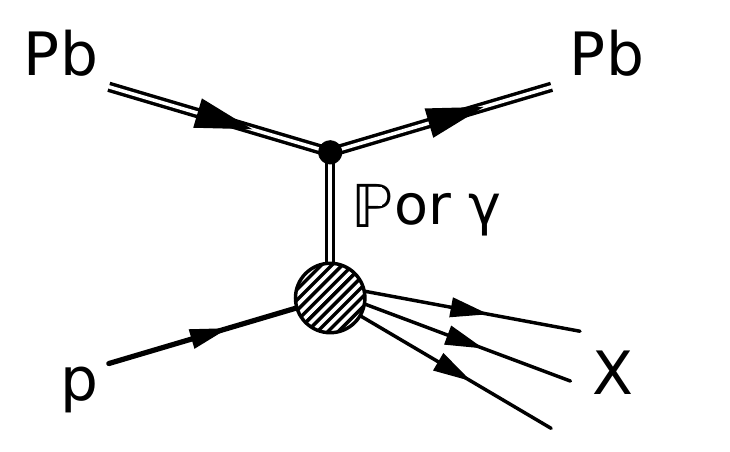}
  \caption{ \label{fig:processes}
    Topologies of \pPb events with large FRG for \PomPb (left) and \PomP or \PhotonP (right).
    The blue and red cones indicate the products of diffractive dissociation for the lead ion and proton, respectively.
    The regions free of final-state particles are marked with green arrows. It is possible for \PhotonPb interactions to mimic the topology on the left but these are highly suppressed compared to the \PhotonP case on the right.
  }
\end{figure*}

Figure~\ref{fig:processes} shows schematic topologies of
single-diffractive pomeron-lead (\PomPb) and pomeron-proton (\PomP) processes for \pPb  collisions.
The HF calorimeters on the side of either lead or proton dissociation are marked with the corresponding color, and are referred to as \HFp or \HFm, depending on the $\eta$ sign.
Single-diffractive dissociation is characterized by a large FRG and an intact proton or ion. Since, for these data, it was not possible to measure the intact protons or ions, the analysis is based on the detection of large FRGs. Double-diffractive dissociation processes, when the proton or ion emitting a pomeron breaks up, result in two sprays of particles separated by a rapidity gap.
If the decay products from the struck proton or ion escape the detector, it is not possible to distinguish such events from single-diffractive ones.

The high Pb nuclear charge, $\PZ_\text{Pb}=82$, enhances the flux of coherent quasi-real photons with respect to the proton by a factor of $\PZ_\text{Pb}^2$. This leads to a significant contribution  of electromagnetic \PhotonP processes  to the sample of events with a large gap on the lead side~\cite{Baltz:2007kq,Khachatryan:2016qhq, Sirunyan:2019nog,Sirunyan:2018sav,Sirunyan:2018fhl}.

\section{Forward rapidity gap distributions}
\label{section:ForwardGapDistributions}
To identify regions devoid of detected final-state particles, the central detector acceptance, $\abs{\eta} < 3$, is divided into 12 $\eta$ bins, each 0.5 units wide, and every bin is considered separately.
The following criteria are used to define ``empty'' bins:

\begin{itemize}
\item For $\abs{\eta} < 2.5$, \ie within the acceptance of the tracker, a given $\eta$ bin is considered to be empty if no
  high purity track~\cite{CMS_TR} with $\pt > 200\MeV$ is found and the total energy of all PF candidates in this bin is less than 6\GeV.
  The use of particle flow objects ensure that bins populated with neutral particles only are not misidentified as being empty.
\item For $2.5 < \abs{\eta} < 3.0$, where no tracking information is available, a bin is considered to be empty if the total energy of all hadronic PF candidates in this bin is less than 13.4\GeV.
  The electronic noise of the ECAL for this region was such that electromagnetic PF candidates were not considered.
\end{itemize}

The energy thresholds are set above the detector noise but still low enough to tag the presence of final-state particles with a good efficiency.
  
The description of  the analysis is based upon the \pPb data.
Each  event in which \HFm has at least one tower with energy greater than 10\GeV is selected and  all empty $\eta$ bins are identified.
If the bin $2.5 < \eta < 3$ is empty, the event is tagged as a \PomPb candidate, as shown in Fig.~\ref{fig:processes} (left), and \dEtaF is defined as the distance from $\eta = 3$ to the nearest edge of the first nonempty $\eta$ bin.
Alternatively, if \HFp has at least one tower above the threshold and the bin  $-3 < \eta < -2.5$ is empty then the event is flagged as a \PomPPhotonP candidate, as is shown in Fig.~\ref{fig:processes} (right). In this case, \dEtaF is the distance from $\eta = -3$ to the nearest edge of the last nonempty $\eta$ bin.
If both \HFp and \HFm are above threshold and FRGs are present on both sides of the detector then the event is counted as both a \PomPb and a \PomPPhotonP candidate.
For \Pbp running, the classification of events is similar but with the  $\eta$ definitions swapped.

The differential cross section of FRG events found in the interval $\abs{\eta} < 3.0$ is defined by
\begin{linenomath*}
  \begin{equation}
    {\left. \dSigmadetaFFrac \right|_{\dEtaF:\abs{\eta} < 3.0}} = {\left. \frac{1}{\Aepsilon\Lumi} \dd{N}{\dEtaF}\right|_{\dEtaF:\abs{\eta} < 3.0}},
    \label{Eqn:CentralSpectrum}
  \end{equation}
\end{linenomath*}
where $N$ is the number of events with  a FRG \dEtaF, \Lumi is the integrated luminosity, $A$ is the detector acceptance, and $\epsilon$ is the efficiency for selecting events.
The product \Aepsilon depends on \dEtaF and is determined for every \dEtaF bin using zero bias data.
The only trigger required for this event sample is the presence of lead and proton bunches crossing the interaction point.
To reject zero bias events without collisions or with only elastic collisions, just those zero bias events
that have at least one high purity track with $\pt > 200\MeV$ were considered.
For a given \dEtaF, \Aepsilon is defined as the fraction of such selected zero bias events that fire the minimum bias trigger, have at least one HF tower with energy exceeding 10\GeV, and have a FRG of size \dEtaF.

The rate of misreconstructed tracks is estimated using non-colliding bunches and is found to be negligible, \ie no fake tracks have been found at the per mille level.
Thus, the track requirement guarantees that the selected zero bias events are indeed inelastic collisions.
The statistical uncertainty in \Aepsilon does not exceed 5\% for large \dEtaF ($\dEtaF \geq 3$) and is negligible for smaller \dEtaF.
The systematic uncertainty caused by the track requirement is estimated using \HIJING and \EPOS MC samples of ZB events, and then comparing the \Aepsilon coefficients obtained with and without the track requirement. The predictions of both MC generators agree rather well and the largest uncertainty of ${\sim}3\%$ is found for bin $1 < \abs{\eta} < 1.5$.

Systematic uncertainties related to the asymmetries in the detector are estimated by comparing the same distributions obtained during the \pPb and \Pbp runs. They vary as a functions of \dEtaF but stay below 13\%.
Utilizing the values of the discrepancy in track reconstruction efficiency between data and simulation found in Ref.~\cite{Khachatryan:2016odn}, the size of that discrepancy in the current study is estimated to be 5\%. This discrepancy induces a systematic uncertainty of less than 5\%. A discrepancy in the PF energy scale is dominated by the HCAL energy scale uncertainty. This has been evaluated in Ref.~\cite{Sirunyan:2019djp} using 40--60\GeV charged hadrons and found to be about 3\%. To account for the significantly lower values of the rapidity gap energy thresholds, and for compatibility with the previous studies of the FRG distributions in pp collisions~\cite{Robert}, the PF energy threshold is varied by 10\% and the full variation in the \dSigmadetaF spectra is taken as an estimate of the corresponding systematic uncertainty.
This uncertainty is about 5\% for the first two \dEtaF bins and smaller for the larger FRG sizes.

The uncertainty in the determination of the integrated luminosity is 3.5\%~\cite{CMS:2018fkg}.
The mean number of inelastic proton-lead collisions per bunch crossing is approximately 0.15. This number decreased slowly during data taking.
The periods with the largest and smallest mean number of simultaneous collisions were used to estimate the sensitivity of the results to collision pileup.
The probability to have pileup in data selected by the MB trigger is evaluated as 10\% and 4\% for these periods, respectively.
The relative difference between the results obtained with the two subsamples is scaled to estimate the influence of pileup on the complete data set. The systematic uncertainty from this effect tends to increase with $\dEtaF$ and reaches a maximum of about 8\% at $\dEtaF = 5.5$.

\begin{figure*}[htb]
  \centering
  \includegraphics[width=1.0\textwidth]{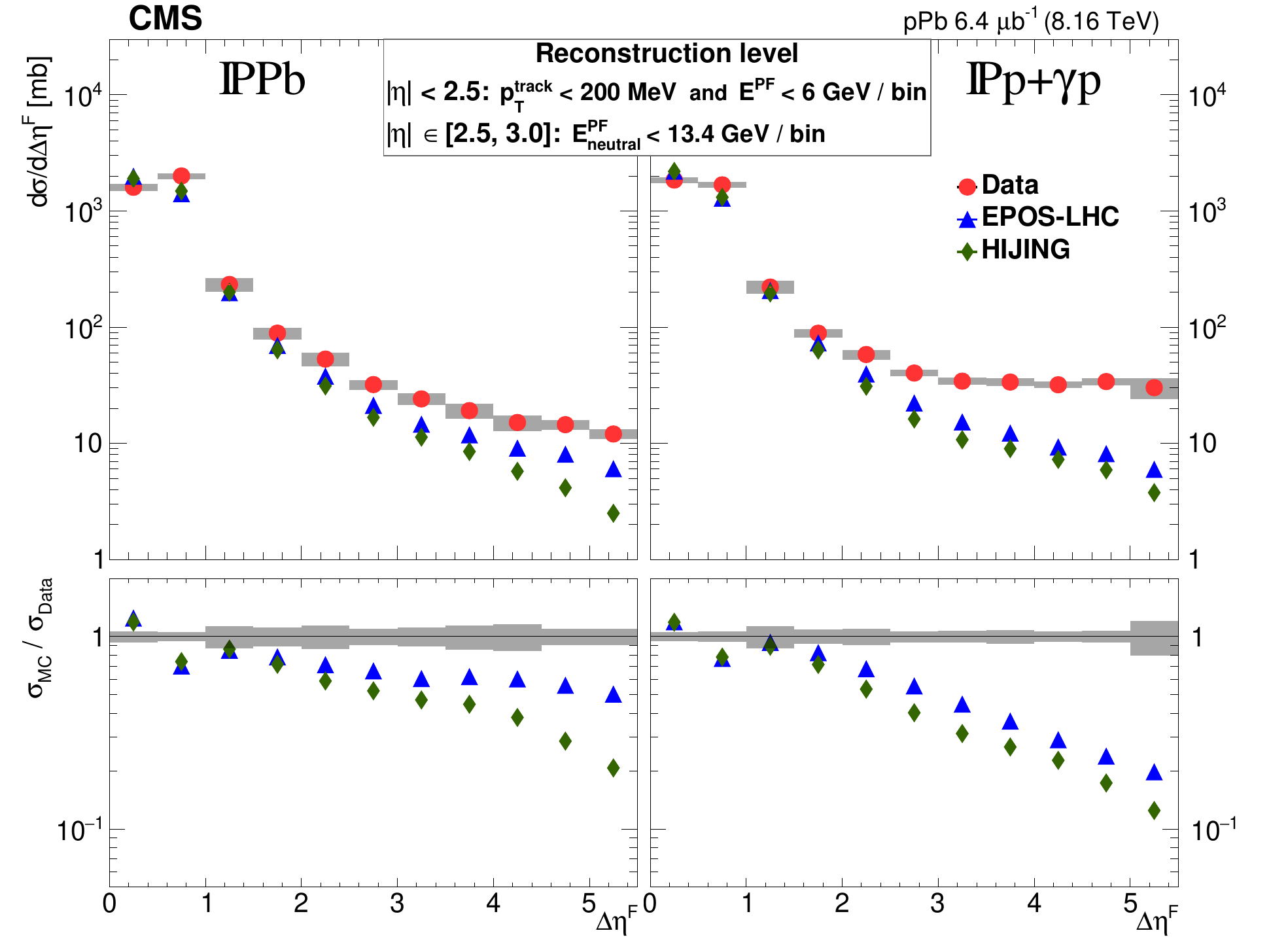}
  \caption{
    Differential cross section \dSigmadetaF for events with \PomPb (left) and \PomPPhotonP (right) topologies obtained at the reconstruction level for $\abs{\eta} < 3.0$ region.  Also shown are the simulated predictions of \EPOS (blue) and \HIJING (green). The statistical and systematic errors are added in quadrature and shown with the gray band. The simulated spectra are normalized to the total visible cross section of the data. The bottom panels show the ratio of simulated predictions to data.
  }
  \label{fig:RG3}
\end{figure*}

Figure~\ref{fig:RG3} shows the reconstruction (detector) level differential cross section \dSigmadetaF for the \PomPb (left) and \PomPPhotonP (right) topologies as well as predictions from the \EPOS  and \HIJING event generators. The results are presented in the laboratory frame of reference. Note that the simulated distributions are normalized to the total visible cross section of the data, $\sigma^\text{vis}=N/\Lumi$, where $N$ is the total number of minimum bias events which have at least one HF tower with energy greater than 10\GeV. The influence of the HF energy scale uncertainty on the normalization factor is evaluated with the MC sample by varying the energy cut by ${\pm}1\GeV$.
This uncertainty is found to be about 1\%. Normalizing the simulated spectra to the total inelastic cross section predicted by the corresponding generators varies the results by less than 20\%.

For both topologies, the spectra fall by a factor of over 50 between \dEtaF = 0 and 2. The low population of the first bin is a consequence of the FRG definition for the regions $2.5 < \abs{\eta} < 3.0$. For $\dEtaF > 2.5$, the spectra tend to flatten off for both topologies.
The predictions of  \EPOS  are closer to the data than those of \HIJING, but neither model gives a good description of the data on the full \dEtaF range.
For the \PomPPhotonP topology, the \EPOS  and \HIJING predictions are significantly below the data in the region $\dEtaF  > 3.$
This suggests that a significant fraction of the events could be from \PhotonP scattering, which is not accounted for in the considered event generators. In the \PomPb case, where the photon flux is a factor $1/{\PZ_\text{Pb}^2}$ smaller, \EPOS  and \HIJING are closer to the data in the region $\dEtaF  > 3.$

\begin{figure*}[htb]
  \centering
  \includegraphics[width=1.0\textwidth]{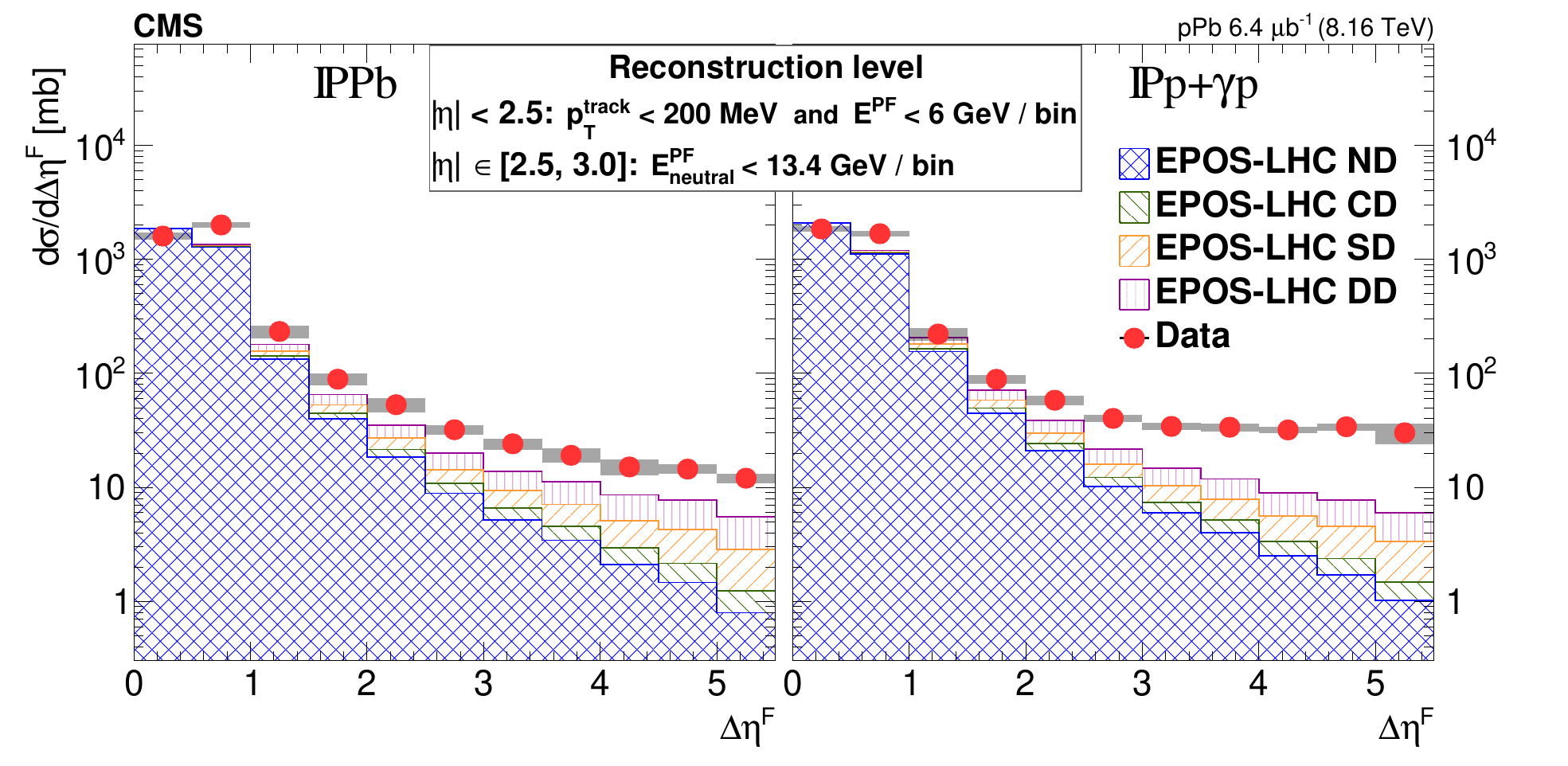}
  \caption{
    Reconstruction level \dSigmadetaF spectra obtained for the central acceptance, $\abs{\eta} < 3$, for the \PomPb (left) and \PomPPhotonP (right) topologies and compared to the corresponding \EPOS predictions.
    The \EPOS predictions are broken down into the nondiffractive (ND, blue), central-diffractive (CD, green), single-diffractive (SD, orange) and double-diffractive (DD, purple) components, shown as stacked contributions.}
  \label{fig:RG3MC}
\end{figure*}

Figure~\ref{fig:RG3MC} shows \EPOS predictions for the  contributions from nondiffractive and diffractive processes to the \dSigmadetaF  spectra. The generator predictions  are broken down into nondiffractive (ND, blue), central-diffractive (CD, green), single-diffractive (SD, orange) and double-diffractive (DD, purple) components, shown as stacked contributions. It is clear that diffractive processes dominate only for large FRGs, \ie $\dEtaF \gtrsim3$.
At the moment, the \HIJING version employed by CMS does not keep the process ID information for generated events, so that it is not possible to disentangle the different physics processes in \HIJING.

\section{Diffraction-enhanced results}

\subsection{Extension of the rapidity gap size acceptance}
\label{subsection:DER}
The sensitivity to diffractive events can be increased by extending the rapidity gap to the region $3.14 < \abs{\eta} < 5.19$ covered by the HF calorimeters.
This was done by reweighting the distributions of Figure~\ref{fig:RG3}, with the probability to have no detectable particles in the HF adjacent to the rapidity gap.
For this study, the HF towers with  $3.00 < \abs{\eta} < 3.14$ are excluded, since they are shadowed by the endcaps of the hadronic calorimeters.
The additional systematic uncertainty introduced by this exclusion is studied with MC samples and found to be negligible.
For each value of \dEtaF, the diffraction-enhanced spectra were calculated by weighting the original $\left. \dSigmadetaF \right|_{\abs{\eta} < 3.0}$ spectra for either the \PomPb or \PomPPhotonP topologies by the probability for the HF calorimeter adjacent to the rapidity gap  to have no towers with a signal distinguishable from noise.

These weights were derived from the data selected by the ZB trigger for each  \dEtaF bin and for each topology.
For every sample of events with at least one high purity track with $\pt > 200\MeV$ and  a certain \dEtaF, the low energy part of the leading tower energy spectrum of the HF calorimeter located adjacent to the rapidity gap is compared to a noise spectrum.
The noise spectra are obtained for \HFp and \HFm individually, using events with noncolliding bunches.
They are then normalized to have the same yield as the  collision spectra for energies within a given range.
The energy range, 1.0--2.9\GeV, is chosen such that the contribution of collision events that have final-state particles in this HF acceptance does not exceed 20\% even for small gap sizes ($\dEtaF < 3$), and is negligible for larger rapidity gaps.
For both the \PomPb and \PomPPhotonP topologies, the probability to have no particles in the corresponding HF increases with \dEtaF because of the decreasing contribution of nondiffractive  events.

This method does not account for the fraction of collision events in which a low momentum particle emitted within the HF acceptance produces a signal that is comparable to the noise level in HF. To evaluate the fraction of such events, the probability to observe an event with such a small signal in the HF adjacent to the rapidity gap is obtained from ZB \EPOS MC samples, using the same approach as for the data. For each rapidity gap size, this probability is compared to the probability to find an event with no stable final-state particles, excluding neutrinos, within the HF acceptance. The relative difference in these fractions is used to correct the obtained results.

For $\dEtaF > 3$, systematic uncertainties from the stability of the normalization procedure and  the dependence of the weights upon  running conditions become dominant. To evaluate systematic uncertainties introduced by both the acceptance correction and the reweighting procedure, the analysis is repeated on zero bias event samples that had at least one high purity track with $\pt > 200\MeV$ to guarantee inelastic collisions.
Events with the HF calorimeter on the side of the rapidity gap having an energy deposition in any tower less then 2\GeV are selected.
The normalized distribution of events with a given \dEtaF, obtained in this way, should be identical within statistical uncertainties to the corresponding distributions obtained from the minimum bias samples after all
corrections and the reweighting.
The relative deviation of the spectra is of order 20\% and is used as an estimate of the systematic uncertainty in the weights.

\subsection{Unfolding}
To correct for detector effects, the diffraction-enhanced spectra are unfolded to the hadron level, that is to the stable final-state particles level. An event at the hadron level is free from the specifics of a given detector and a direct comparison of unfolded data and MC generator predictions can be made. For the purposes of the unfolding, empty $\eta$ bins, each 0.5 $\eta$ units wide, within the region $\abs{\eta} < 2.5$ are defined as having no charged particles with $\pt$ above $200\MeV$ and as having the total energy of all detectable particles below 6\GeV.
Here and later, all stable particles other than neutrinos are considered to be detectable.
For the edge bins, $2.5 < \abs{\eta} < 3.0$, this requirement is changed such that the  total energy of neutral hadrons does not exceed 13.4\GeV.
For the HF acceptance, $3.14 < \abs{\eta} < 5.19$, a rapidity gap is defined as an absence of any detectable particles.
The unfolding is performed using the D'Agostini iteration method with early stopping~\cite{DAgostini:1994fjx} implemented in \textsc{r}oo\textsc{u}nfold~\cite{Adye:2011gm}. The optimal number of iterations is chosen according to the minimum
of the average global correlation coefficient~\cite{Schmitt:UnfoldingMethods} and is equal to 2. Response matrices are constructed for \Pbp and \pPb data samples separately, for both \PomPb and the \PomPPhotonP topologies using \EPOS generator.

\begin{figure*}[htb!]
  \centering
  \includegraphics[width=1.0\textwidth]{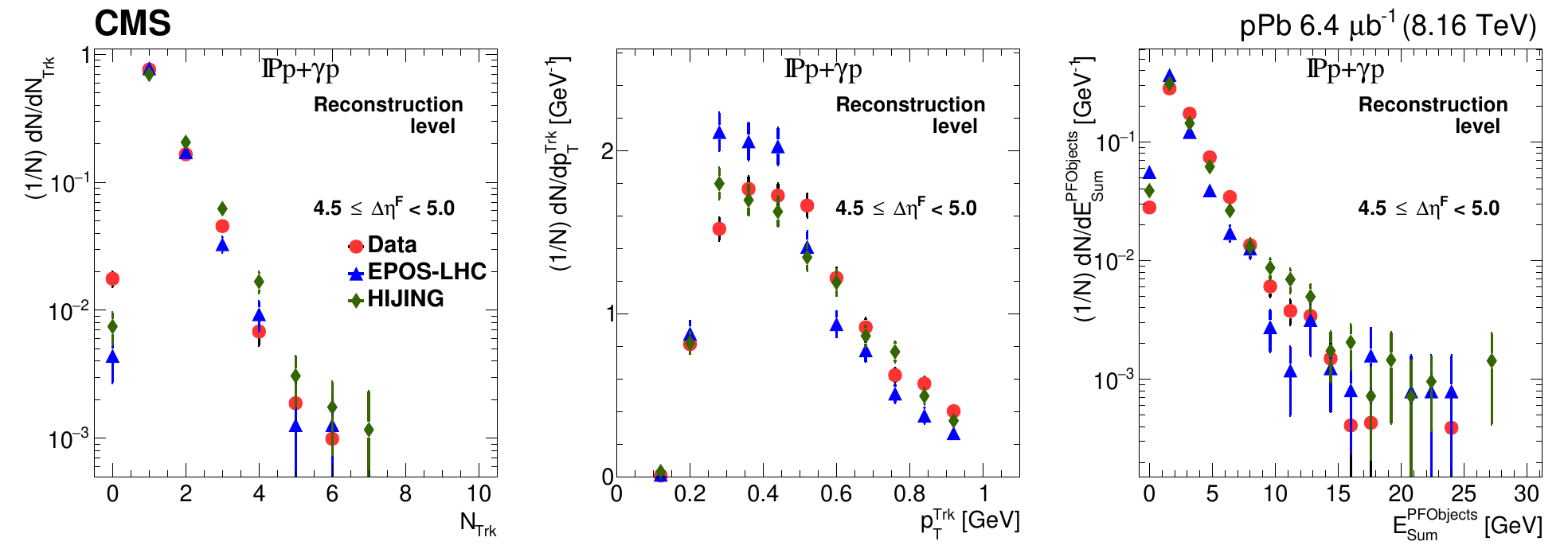}
  \caption{
    The number of high purity tracks, $N_{\mathrm{Trk}}$ (left), their transverse momentum, \pt, (middle), and the total energy of all PF candidates, $E^{\text{PFObjects}}_{\text{Sum}}$, (right) in the first $\eta$ bin after a gap of $4.5 < \dEtaF < 5.0$, for events with the \PomPPhotonP topology. Also shown are the corresponding distributions for the \EPOS and \HIJING generators.
  }
  \label{fig:spectra}
\end{figure*}

The right hand panel of Fig.~\ref{fig:RG3} suggests that the \PomPPhotonP data set contains a large contribution from \PhotonP events.  Neither the \EPOS nor \HIJING generators account for such events. To test the appropriateness of using these generators for the unfolding in this case, the three variables used for the hadron definition of \dEtaF, namely the number of tracks, their \pt distributions, and the sum of the energy of all PF candidates in a given bin are studied.
For each \dEtaF bin, the distributions of these variables in the first nonempty $\eta$ bin are compared in data and simulation.
Figure~\ref{fig:spectra} shows an example of those distributions at the reconstruction level for diffraction-enhanced events with the \PomPPhotonP topology for $4.5\leqslant\dEtaF < 5$.
Both the \EPOS and \HIJING predictions for the first nonempty bins are found to be in good agreement with the data.

The stability of the unfolding procedure is studied with respect to the model dependence of the response matrices, imperfections in detector response simulation, and finite MC statistics.
To account for the physics model dependence, a closure test is performed by unfolding the spectrum of reconstructed \HIJING samples using the \EPOS response matrices. The nonclosure is statistically significant only for $\dEtaF < 3$. In this region the nonclosure did not exceed 11\%.
The magnitude of the nonclosure is used as the estimate of the  corresponding systematic uncertainty.
To study the stability of the unfolding procedure with respect to imperfections in the simulation of  detector performance, the influence of the track reconstruction efficiency and PF energy thresholds were studied as
in Section~\ref{section:ForwardGapDistributions} above. The track reconstruction efficiency is varied by 5\% and the PF energy threshold, at the hadron level, is varied by 10\%. The difference after smoothing in the spectra is used to evaluate the systematic uncertainty.
The resulting combined uncertainties were found to be about 10\% for small \dEtaF ($\dEtaF < 3$), decreasing towards \dEtaF =  3. Systematic uncertainties related to the limited counts in the MC samples used to construct the response matrices are found to be below 1\%.

\subsection{Results}
\begin{figure*}[htbp!]
  \centering
  \includegraphics[width=1.0\textwidth]{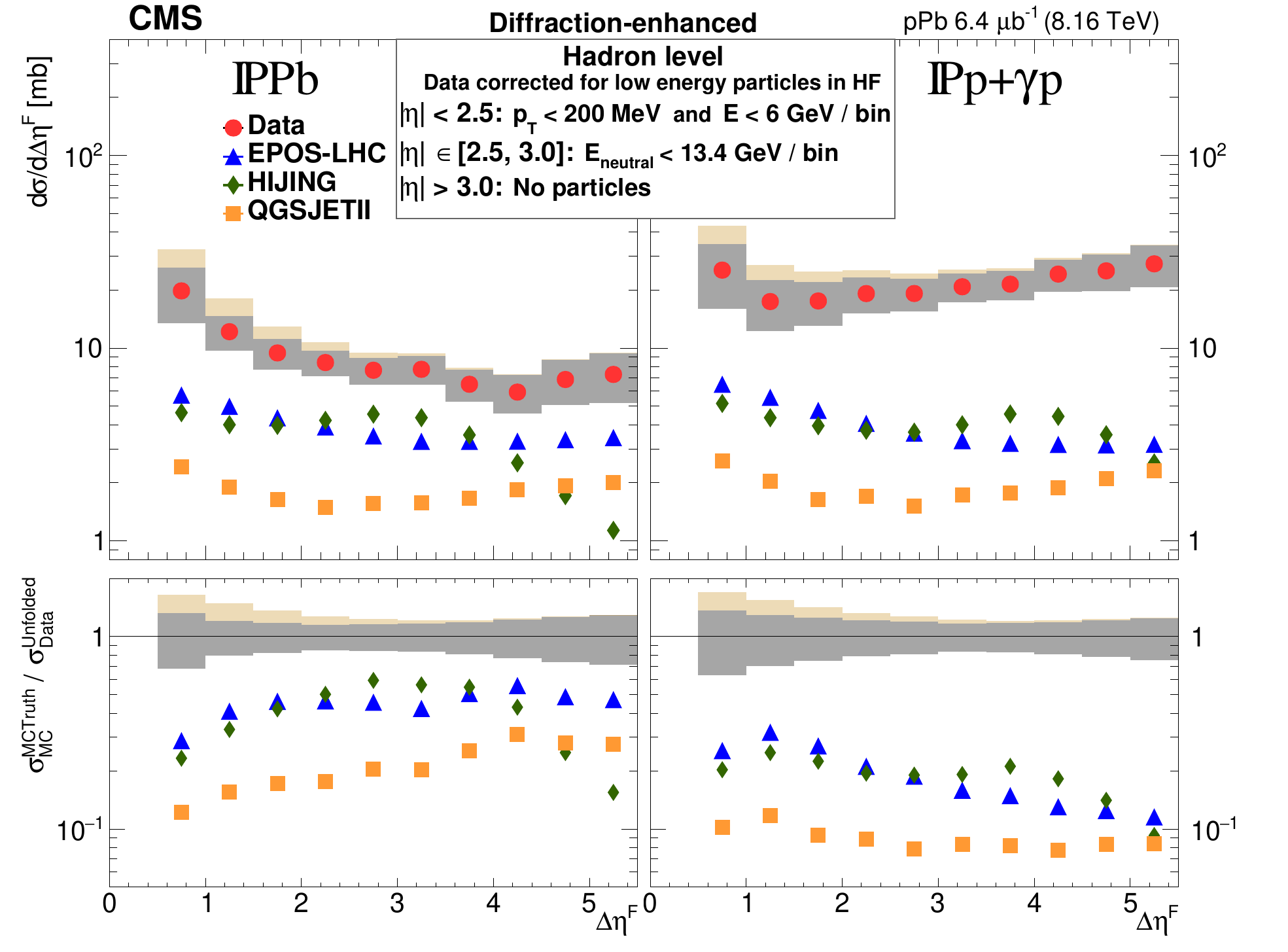}
  \caption{
    Unfolded diffraction-enhanced differential cross section \dSigmadetaF spectra compared to hadron level predictions of the  \EPOS, \HIJING, and \QGSJET generators.
    The data are corrected for the contribution from events with undetectable energy in the HF calorimeter adjacent to the FRG. The corrections are obtained using the \EPOS MC samples.
    For the \pPb data sample, in the \PomPb case (left), the FRG, \dEtaF, is measured from $\eta = 3$ and no particles are present within $3.00 < \eta < 5.19$, while for the \PomPPhotonP case (right), the FRG is measured from $\eta = -3$ and no particles are present within $-5.19 < \eta < -3.00$.
    The  statistical and systematic uncertainties are added in quadrature. The gray band shows the resulting uncertainty.
    The yellow band indicates the values of the only MC-based correction done to account for the HF calorimeter energy deposition below the noise level. The bottom panels show the ratio of the predictions of the three generators to data.
  }
  \label{fig:unfoldedwithsysDataCorrected}
\end{figure*}

The diffraction-enhanced \dEtaF  distributions are obtained as a weighted mean of the  \pPb and \Pbp spectra unfolded with \EPOS, with weights defined by the statistical uncertainties in the two spectra. The spectra are shown in Fig.~\ref{fig:unfoldedwithsysDataCorrected}, together with hadron level predictions from the \EPOS, \QGSJET and \HIJING generators. The results are presented in the laboratory frame of reference. The increase of the rapidity gap is implemented statistically as described in Sec.~\ref{subsection:DER}, and is equivalent to a requirement of no final-state particles within the corresponding HF acceptance, $3.00 < \eta < 5.19$ for the \PomPb case, or $-5.19 < \eta < -3.00$ for the \PomPPhotonP case. So, although the abscissa axis represents the rapidity gap size seen in the central detector acceptance, $\abs{\eta} < 3.0$, the actual gap size is 2.19 units larger for the diffraction-enhanced distributions. The results for $0 < \dEtaF < 0.5$ are not included in the plots because of the poor rapidity gap sensitivity discussed in Section~\ref{section:ForwardGapDistributions}.

All predicted cross sections of the generators are below the data for both the \PomPb and \PomPPhotonP cases. For the \PomPPhotonP topology, the data are a factor of at least 5 above the generator predictions, suggesting a strong contribution from \PhotonP events. As for the \PomPb case, despite the discrepancy in magnitude between \PomPb data and MC predictions, the  shape of the data spectrum is  quite well described  by the \EPOS and \QGSJET generators. However, \HIJING demonstrates a sharp decline at large \dEtaF in contrast to the data. At large \dEtaF, there is a slight rise in the data as well as in the cosmic ray MC generator, \QGSJET. The excess of the data over each of the  MC predictions is  largest at low \dEtaF.

Both the ATLAS and CMS Collaborations have measured the differential cross section for rapidity gap events for 7\TeV pp collisions~\cite{Robert, ATLASFRG}. Although these spectra were measured at a different collision energy and with a rapidity gap definition different from the one used in this analysis, a rough comparison with current results is still possible. The ratio of the \PomPb and pp differential cross sections at large \dEtaF can be explained by a factor $A^{\alpha}$, where $\alpha \simeq 1/3$, as expected in the Glauber--Gribov approach.

\begin{figure*}[htb!]
  \centering
  \includegraphics[width=1.0\textwidth]{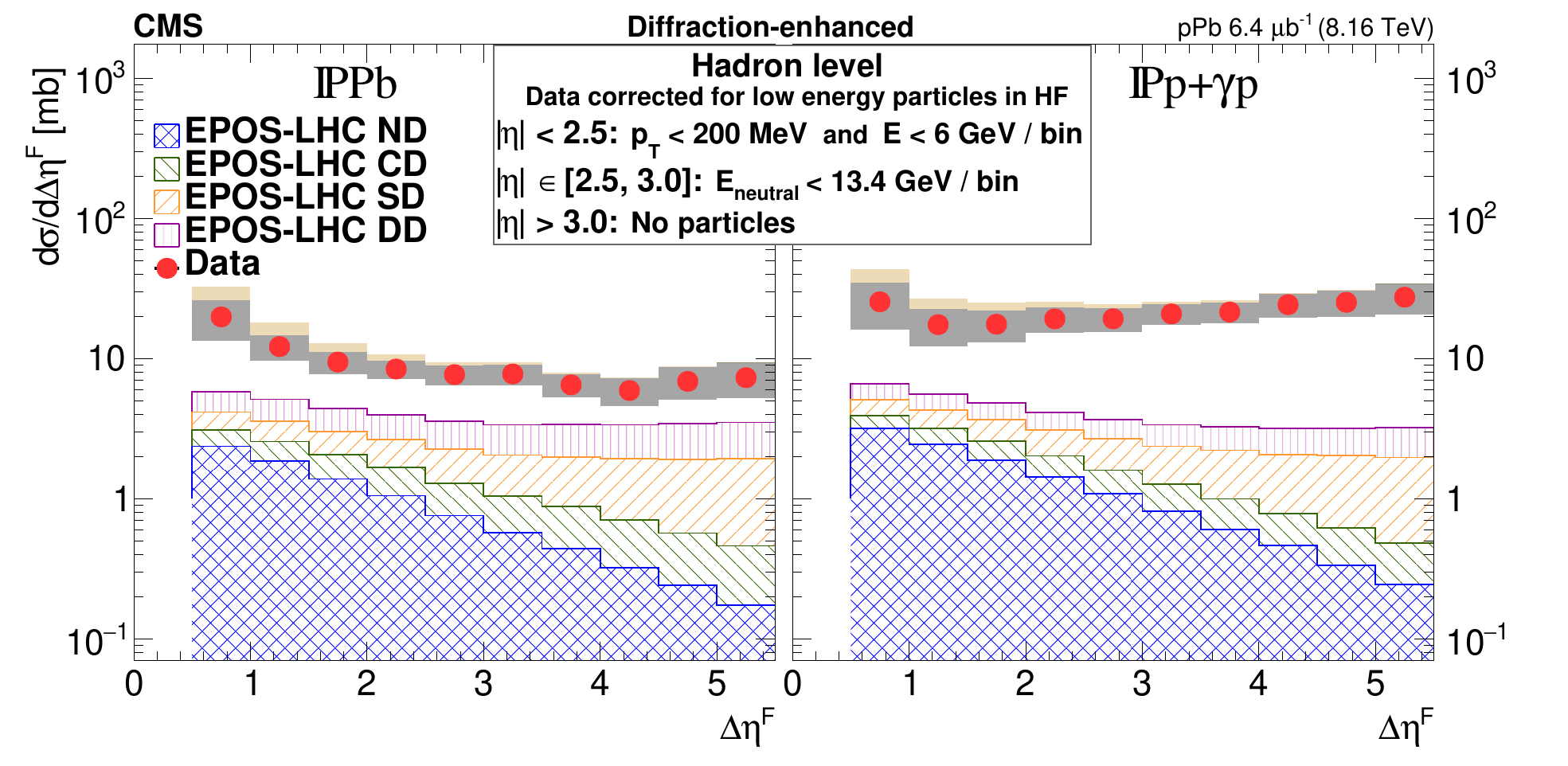}
  \caption{Unfolded diffraction-enhanced differential cross section \dSigmadetaF spectra for the \PomPb (left) and \PomPPhotonP (right) topologies compared to the \EPOS predictions. The \EPOS predictions are broken down into the nondiffractive (ND, blue), central-diffractive (CD, green), single-diffractive (SD, orange) and double-diffractive (DD, purple) components, shown as stacked contributions.}
  \label{fig:EposRG5}
\end{figure*}

\begin{figure*}[htb!]
  \centering
  \includegraphics[width=1.0\textwidth]{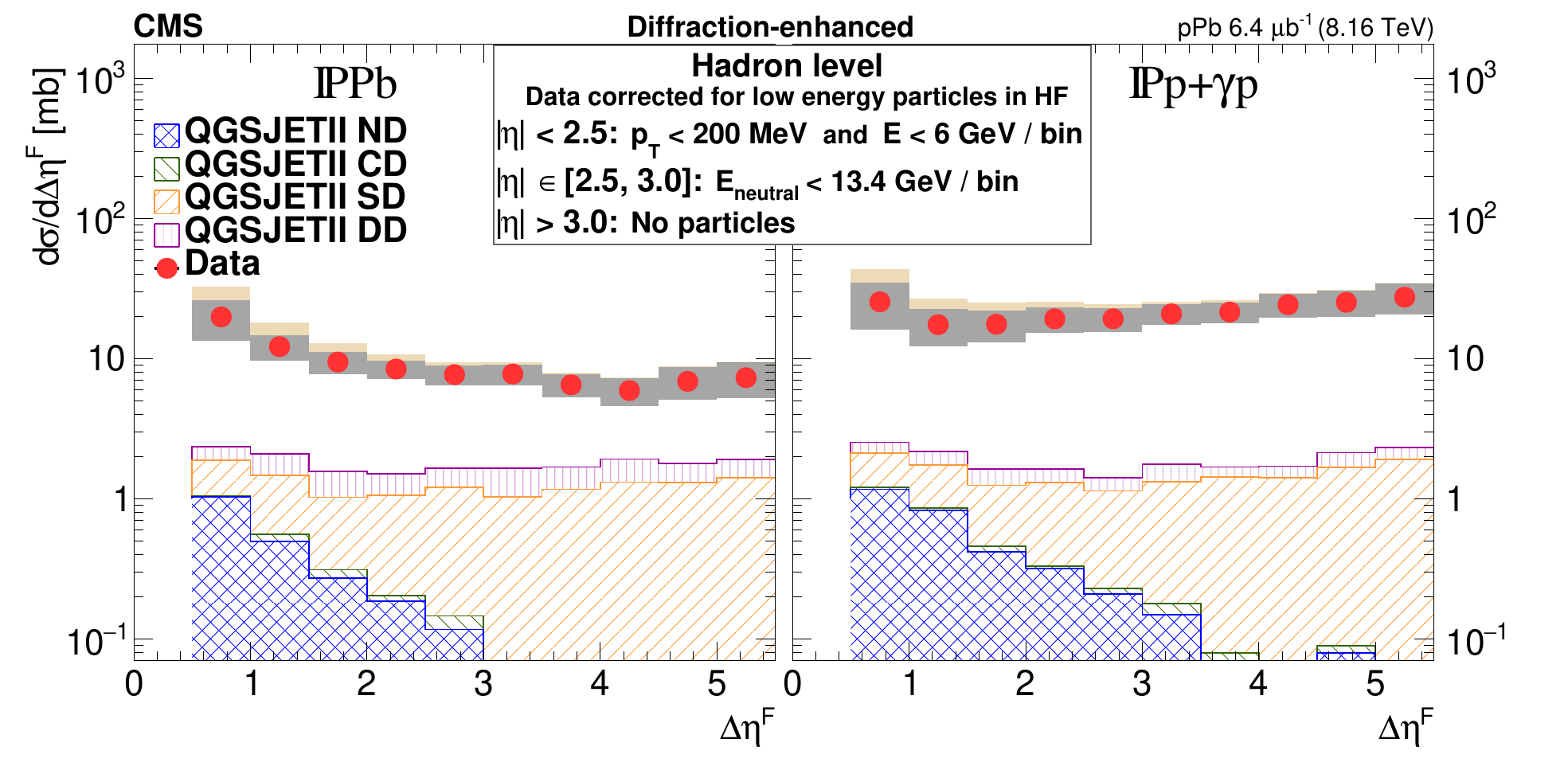}
  \caption{Unfolded diffraction-enhanced differential cross section \dSigmadetaF spectra for the \PomPb (left) and \PomPPhotonP (right) topologies compared to the \QGSJET predictions. The \QGSJET predictions are broken down into the nondiffractive (ND, blue), central-diffractive (CD, green), single-diffractive (SD, orange) and double-diffractive (DD, purple) components, shown as stacked contributions.}
  \label{fig:qgsjetRG5}
\end{figure*}

Figures~\ref{fig:EposRG5} and \ref{fig:qgsjetRG5} show the unfolded diffraction-enhanced \dSigmadetaF spectra for the \PomPb (left) and \PomPPhotonP (right) topologies compared to the detailed predictions of the \EPOS and \QGSJET generators. The generator predictions are broken down into nondiffractive (ND, blue), central-diffractive (CD, green), single-diffractive (SD, orange) and double-diffractive (DD, purple) components, shown as stacked contributions.

\subsection{Data driven cross check}
To provide a data driven cross check of the MC-based correction to the events in which low momentum particles are emitted within the HF acceptance, a complimentary analysis of the diffraction-enhanced data sample has been performed. The forward instrumentation of CMS, the Zero-Degree Calorimeters (ZDCs), can detect the lead break-up products, so events with an intact lead can be selected explicitly for the \PomPPhotonP topology. For those events the probability to have any final state particles in the HF acceptance at the side of the escaping ion is significantly smaller. Such a comparison provides a good cross check to the main analysis approach.

The ZDCs are located 140\unit{m} away from the CMS interaction point and consist of tungsten absorber and quartz fibers. Being located behind the LHC dipole magnets, the ZDC calorimeters are sensitive to neutral particles produced in collisions at high pseudorapidity $\abs{\eta} > 8.3$~\cite{Suranyi:2019mgm,Suranyi:2021ssd} while charged particles are swept away by the dipole.
The ZDC located on the negative side of the CMS detector (\ZDCm) allows for the exclusion of events with strongly boosted energetic neutrons produced from the break up of the lead nucleus. A subset of events with an intact lead nucleus can be selected requiring \ZDCm energy to be below 1\TeV~\cite{Suranyi:2019mgm,Suranyi:2021ssd}.

\begin{figure}[htb!]
  \centering
  \includegraphics[width=0.5\textwidth]{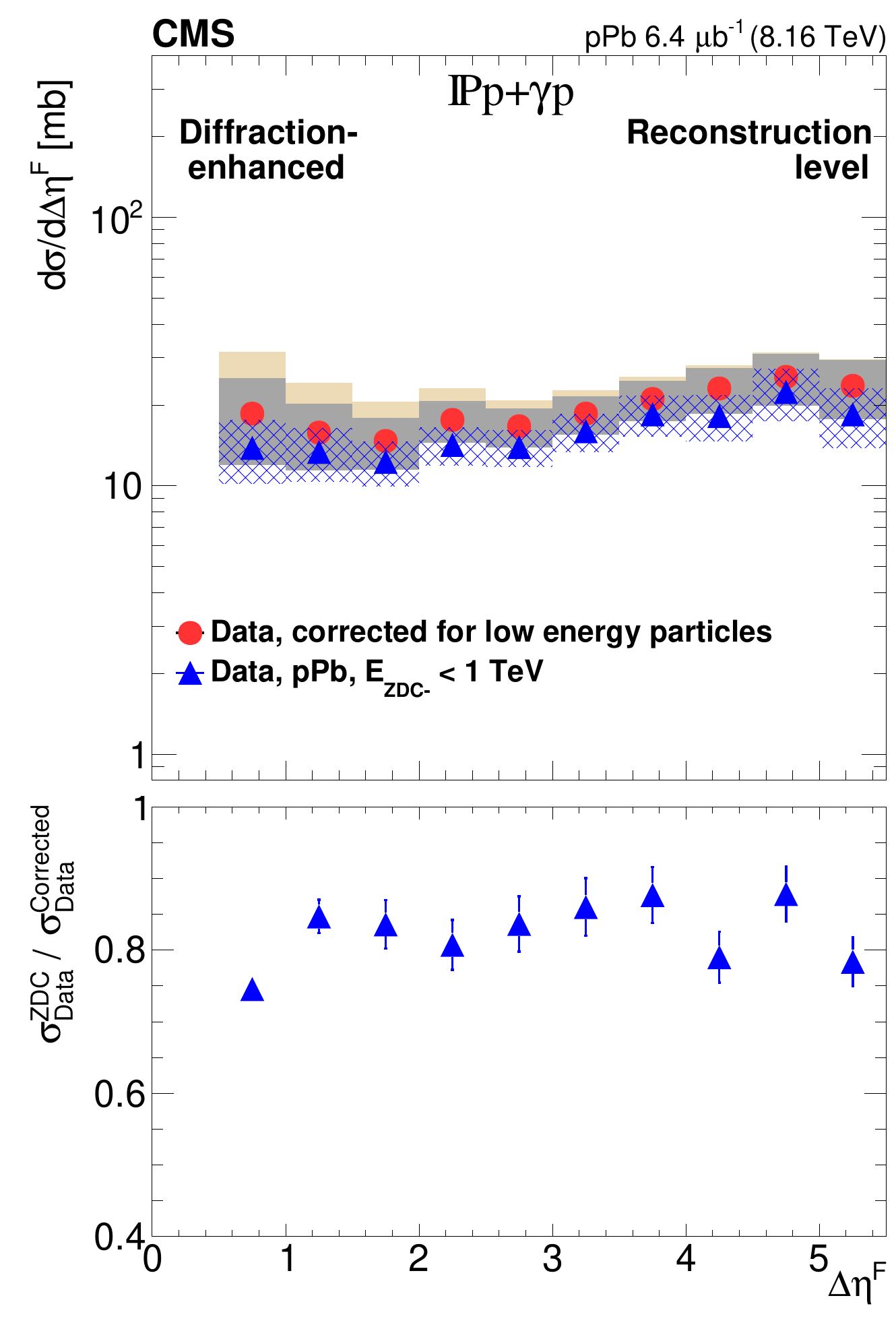}
  \caption{
    Top: Reconstruction level diffraction-enhanced differential cross section \dSigmadetaF spectrum corrected for the contribution from events with undetectable energy in the HF calorimeter adjacent to the rapidity gap. The correction value is indicated with the yellow band. The statistical and systematic uncertainties are added in quadrature. The gray band shows the resulting uncertainty. The distribution is shown together with the spectrum obtained with events satisfying the ZDC veto requirement $E_{\ZDCm} < 1\TeV$ which selects only the events without lead nuclear break up. No correction for HF undetectable energy is applied to this distribution and no systematic uncertainties related to the ZDC veto are accounted for. The statistical and systematic uncertainties are added in quadrature.
    Bottom: A fraction of events selected with the ZDC veto requirement as a function of the rapidity gap size. Only statistical uncertainties are shown in the plot.
  }
  \label{fig:ZDCm}
\end{figure}

Figure~\ref{fig:ZDCm},~top, compares the reconstruction level diffraction-enhanced \dSigmadetaF spectrum obtained for the \PomPPhotonP event topology and corrected for the contribution of the low momentum particles in the HF acceptance to the corresponding distribution for the events passing the \ZDCm veto requirement.
Within the systematic uncertainties the diffraction-enhanced results agree with the results obtained with the ZDC veto.
Figure~\ref{fig:ZDCm},~bottom, shows the fraction of events selected with the ZDC veto requirement as a function of the rapidity gap size.
The ratio is flat as a function of \dEtaF.

\clearpage
\section{Summary}
For the first time at the CERN LHC, the forward rapidity gap spectra \dSigmadetaF for proton-lead (\pPb) collisions at a center-of-mass energy per nucleon pair of $\sqrtsNN = 8.16\TeV$ have been measured for both pomeron-lead (\PomPb) and pomeron-proton (\PomPPhotonP) topologies.
For the latter topology, predictions from the \EPOS, \QGSJET, and \HIJING generators are a factor of five or more below the data for large rapidity gaps.
This suggests a very strong contribution from \PhotonP interactions, which are not included in these event generators.
For the \PomPb topology, where the photon-exchange contribution is expected to be negligible, the \EPOS predictions are about a factor of 2 and the \QGSJET predictions are about a factor of 4 below the data.
However, the shape of the \dSigmadetaF spectrum is similar to that in data for both these generators.
In contrast to the data, \HIJING demonstrates a sharp decline for large rapidity gaps.
These results will be helpful to understand the high energy limit of quantum chromodynamics and in modeling cosmic ray air showers.

\begin{acknowledgments}
We congratulate our colleagues in the CERN accelerator departments for the excellent performance of the LHC and thank the technical and administrative staffs at CERN and at other CMS institutes for their contributions to the success of the CMS effort. In addition, we gratefully acknowledge the computing centers and personnel of the Worldwide LHC Computing Grid and other centers for delivering so effectively the computing infrastructure essential to our analyses. Finally, we acknowledge the enduring support for the construction and operation of the LHC, the CMS detector, and the supporting computing infrastructure provided by the following funding agencies: BMBWF and FWF (Austria); FNRS and FWO (Belgium); CNPq, CAPES, FAPERJ, FAPERGS, and FAPESP (Brazil); MES and BNSF (Bulgaria); CERN; CAS, MoST, and NSFC (China); MINCIENCIAS (Colombia); MSES and CSF (Croatia); RIF (Cyprus); SENESCYT (Ecuador); MoER, ERC PUT and ERDF (Estonia); Academy of Finland, MEC, and HIP (Finland); CEA and CNRS/IN2P3 (France); BMBF, DFG, and HGF (Germany); GSRI (Greece); NKFIH (Hungary); DAE and DST (India); IPM (Iran); SFI (Ireland); INFN (Italy); MSIP and NRF (Republic of Korea); MES (Latvia); LAS (Lithuania); MOE and UM (Malaysia); BUAP, CINVESTAV, CONACYT, LNS, SEP, and UASLP-FAI (Mexico); MOS (Montenegro); MBIE (New Zealand); PAEC (Pakistan); MES and NSC (Poland); FCT (Portugal); MESTD (Serbia); MCIN/AEI and PCTI (Spain); MOSTR (Sri Lanka); Swiss Funding Agencies (Switzerland); MST (Taipei); MHESI and NSTDA (Thailand); TUBITAK and TENMAK (Turkey); NASU (Ukraine); STFC (United Kingdom); DOE and NSF (USA).

\hyphenation{Rachada-pisek} Individuals have received support from the Marie-Curie program and the European Research Council and Horizon 2020 Grant, contract Nos.\ 675440, 724704, 752730, 758316, 765710, 824093, 884104, and COST Action CA16108 (European Union); the Leventis Foundation; the Alfred P.\ Sloan Foundation; the Alexander von Humboldt Foundation; the Belgian Federal Science Policy Office; the Fonds pour la Formation \`a la Recherche dans l'Industrie et dans l'Agriculture (FRIA-Belgium); the Agentschap voor Innovatie door Wetenschap en Technologie (IWT-Belgium); the F.R.S.-FNRS and FWO (Belgium) under the ``Excellence of Science -- EOS" -- be.h project n.\ 30820817; the Beijing Municipal Science \& Technology Commission, No. Z191100007219010; the Ministry of Education, Youth and Sports (MEYS) of the Czech Republic; the Hellenic Foundation for Research and Innovation (HFRI), Project Number 2288 (Greece); the Deutsche Forschungsgemeinschaft (DFG), under Germany's Excellence Strategy -- EXC 2121 ``Quantum Universe" -- 390833306, and under project number 400140256 - GRK2497; the Hungarian Academy of Sciences, the New National Excellence Program - \'UNKP, the NKFIH research grants K 124845, K 124850, K 128713, K 128786, K 129058, K 131991, K 133046, K 138136, K 143460, K 143477, 2020-2.2.1-ED-2021-00181, and TKP2021-NKTA-64 (Hungary); the Council of Science and Industrial Research, India; the Latvian Council of Science; the Ministry of Education and Science, project no. 2022/WK/14, and the National Science Center, contracts Opus 2021/41/B/ST2/01369 and 2021/43/B/ST2/01552 (Poland); the Funda\c{c}\~ao para a Ci\^encia e a Tecnologia, grant CEECIND/01334/2018 (Portugal); the National Priorities Research Program by Qatar National Research Fund; MCIN/AEI/10.13039/501100011033, ERDF ``a way of making Europe", and the Programa Estatal de Fomento de la Investigaci{\'o}n Cient{\'i}fica y T{\'e}cnica de Excelencia Mar\'{\i}a de Maeztu, grant MDM-2017-0765 and Programa Severo Ochoa del Principado de Asturias (Spain); the Chulalongkorn Academic into Its 2nd Century Project Advancement Project, and the National Science, Research and Innovation Fund via the Program Management Unit for Human Resources \& Institutional Development, Research and Innovation, grant B05F650021 (Thailand); the Kavli Foundation; the Nvidia Corporation; the SuperMicro Corporation; the Welch Foundation, contract C-1845; and the Weston Havens Foundation (USA).
\end{acknowledgments}

\bibliography{auto_generated}

\providecommand{\href}[2]{#2}\begingroup\raggedright\begin{thebibliography}{10}%
\makeatletter
\providecommand{\hrefCMSnoop }[0]{\@secondoftwo}%
\makeatother
\providecommand{\doi}{\texttt{doi:}\begingroup \urlstyle{tt}\Url}

\bibitem{Gribov:1961ex}
\hrefCMSnoop {}{V.~N. Gribov, ``Possible asymptotic behavior of elastic
  scattering'',} \textit{ JETP Lett.} \textbf{ 41} (1961) 667.

\bibitem{Chew:1961ev}
\hrefCMSnoop {}{G.~F. Chew and S.~C. Frautschi, ``Principle of equivalence for
  all strongly interacting particles within the {S}-matrix framework'',}
  \textit{ Phys. Rev. Lett.} \textbf{ 7} (1961) 394,
  \href{http://dx.doi.org/10.1103/PhysRevLett.7.394}{\doi{10.1103/PhysRevLett.7.394}}.

\bibitem{Low:1975sv}
\hrefCMSnoop {}{F.~E. Low, ``A model of the bare pomeron'',} \textit{ Phys.
  Rev. D} \textbf{ 12} (1975) 163,
  \href{http://dx.doi.org/10.1103/PhysRevD.12.163}{\doi{10.1103/PhysRevD.12.163}}.

\bibitem{Nussinov:1975mw}
\hrefCMSnoop {}{S.~Nussinov, ``Colored quark version of some hadronic
  puzzles'',} \textit{ Phys. Rev. Lett.} \textbf{ 34} (1975) 1286,
  \href{http://dx.doi.org/10.1103/PhysRevLett.34.1286}{\doi{10.1103/PhysRevLett.34.1286}}.

\bibitem{Fadin:1975cb}
\hrefCMSnoop {}{V.~S. Fadin, E.~A. Kuraev, and L.~N. Lipatov, ``On the
  {Pomeranchuk} singularity in asymptotically free theories'',} \textit{ Phys.
  Lett. B} \textbf{ 60} (1975) 50,
  \href{http://dx.doi.org/10.1016/0370-2693(75)90524-9}{\doi{10.1016/0370-2693(75)90524-9}}.

\bibitem{Kaidalov:1979jz}
\hrefCMSnoop {}{A.~B. Kaidalov, ``Diffractive production mechanisms'',}
  \textit{ Phys. Rept.} \textbf{ 50} (1979) 157,
  \href{http://dx.doi.org/10.1016/0370-1573(79)90043-7}{\doi{10.1016/0370-1573(79)90043-7}}.

\bibitem{Donnachie:1984xq}
\hrefCMSnoop {}{A.~Donnachie and P.~V. Landshoff, ``Elastic scattering and
  diffraction dissociation'',} \textit{ Nucl. Phys. B} \textbf{ 244} (1984)
  322,
  \href{http://dx.doi.org/10.1016/0550-3213(84)90315-8}{\doi{10.1016/0550-3213(84)90315-8}}.

\bibitem{N.Cartiglia:2015gve}
\hrefCMSnoop {}{{{LHC} {Forward Physics} Working Group} Collaboration, ``{LHC}
  {Forward Physics}'',} \textit{ J. Phys. G} \textbf{ 43} (2016) 110201,
  \href{http://dx.doi.org/10.1088/0954-3899/43/11/110201}{\doi{10.1088/0954-3899/43/11/110201}},
  \href{http://www.arXiv.org/abs/1611.05079}{\texttt{arXiv:1611.05079}}.

\bibitem{Gribov:1968jf}
\hrefCMSnoop {}{V.~N. Gribov, ``{Glauber} corrections and the interaction
  between high-energy hadrons and nuclei'',} \textit{ Sov. Phys. JETP} \textbf{
  29} (1969) 483.

\bibitem{Glauber:1959}
\hrefCMSnoop {}{R.~J. Glauber, ``High-energy collision theory'',} in \textit{
  Lectures in theoretical physics}, W.~E. Brittin and L.~G. Dunham, eds.,
  volume~1, p.~315.
\newblock Interscience Publishers, New York, 1959.

\bibitem{Kopeliovich:1981pz}
\href {http://jetpletters.ru/ps/1511/article_23095.shtml}{B.~Z. Kopeliovich,
  L.~I. Lapidus, and A.~B. Zamolodchikov, ``Dynamics of color in hadron
  diffraction on nuclei'',} \textit{ JETP Lett.} \textbf{ 33} (1981) 595.

\bibitem{x15}
\hrefCMSnoop {}{B.~Z. Kopeliovich, I.~K. Potashnikova, and I.~Schmidt, ``Large
  rapidity gap processes in proton-nucleus collisions'',} \textit{ Phys. Rev.
  C} \textbf{ 73} (2006) 034901,
  \href{http://dx.doi.org/10.1103/PhysRevC.73.034901}{\doi{10.1103/PhysRevC.73.034901}},
  \href{http://www.arXiv.org/abs/hep-ph/0508277}{\texttt{arXiv:hep-ph/0508277}}.

\bibitem{HELIOS}
\hrefCMSnoop {}{{{HELIOS}} Collaboration, ``Diffraction dissociation of nuclei
  in 450 {GeV}$/c$ proton-nucleus collisions'',} \textit{ Z. Phys. C} \textbf{
  49} (1991) 355,
  \href{http://dx.doi.org/10.1007/BF01549687}{\doi{10.1007/BF01549687}}.

\bibitem{EHS}
\hrefCMSnoop {}{{{EHS/NA22}} Collaboration, ``Reactions with leading hadrons in
  meson-proton interactions at 250 {GeV}$/c$'',} \textit{ Z. Phys. C} \textbf{
  75} (1997) 229,
  \href{http://dx.doi.org/10.1007/s002880050465}{\doi{10.1007/s002880050465}}.

\bibitem{x14}
\hrefCMSnoop {}{A.~B. Kaidalov, V.~A. Khoze, A.~D. Martin, and M.~G. Ryskin,
  ``Diffraction of protons and nuclei at high-energies'',} \textit{ Acta Phys.
  Polon. B} \textbf{ 34} (2003) 3163,
  \href{http://www.arXiv.org/abs/hep-ph/0303111}{\texttt{arXiv:hep-ph/0303111}}.

\bibitem{x16}
\hrefCMSnoop {}{L.~Frankfurt and M.~Strikman, ``Novel {QCD} phenomena in {pA}
  collisions at {LHC}'',} in \textit{ 2nd Workshop on Hard Probes in Heavy Ion
  Collisions at the LHC: 2nd Plenary Meeting Geneva, Switzerland, March 11-15,
  2002}.
\newblock CERN, Geneva, Switzerland, 2002.
\newblock
  \href{http://www.arXiv.org/abs/hep-ph/0210088}{\texttt{arXiv:hep-ph/0210088}}.

\bibitem{x17}
\hrefCMSnoop {}{V.~Guzey and M.~Strikman, ``Proton-nucleus scattering and cross
  section fluctuations at {RHIC} and {LHC}'',} \textit{ Phys. Lett. B} \textbf{
  633} (2006) 245,
  \href{http://dx.doi.org/10.1016/j.physletb.2005.11.065}{\doi{10.1016/j.physletb.2005.11.065}},
  \href{http://www.arXiv.org/abs/hep-ph/0505088}{\texttt{arXiv:hep-ph/0505088}}.

\bibitem{Deng:2010mv}
\hrefCMSnoop {}{W.-T. Deng, X.-N. Wang, and R.~Xu, ``Hadron production in
  {p+p}, {p+Pb}, and {Pb+Pb} collisions with the {HIJING} 2.0 model at energies
  available at the {CERN} {Large Hadron Collider}'',} \textit{ Phys. Rev. C}
  \textbf{ 83} (2011) 014915,
  \href{http://dx.doi.org/10.1103/PhysRevC.83.014915}{\doi{10.1103/PhysRevC.83.014915}},
  \href{http://www.arXiv.org/abs/1008.1841}{\texttt{arXiv:1008.1841}}.

\bibitem{Pierog:2013ria}
T.~Pierog\hrefCMSnoop {}{ {et~al.}, ``{EPOS LHC}: Test of collective
  hadronization with data measured at the {CERN} {Large Hadron Collider}'',}
  \textit{ Phys. Rev. C} \textbf{ 92} (2015) 034906,
  \href{http://dx.doi.org/10.1103/PhysRevC.92.034906}{\doi{10.1103/PhysRevC.92.034906}},
  \href{http://www.arXiv.org/abs/1306.0121}{\texttt{arXiv:1306.0121}}.

\bibitem{Ostapchenko:2010vb}
\hrefCMSnoop {}{S.~Ostapchenko, ``{Monte Carlo} treatment of hadronic
  interactions in enhanced {Pomeron} scheme: I. {QGSJET-II} model'',} \textit{
  Phys. Rev. D} \textbf{ 83} (2011) 014018,
  \href{http://dx.doi.org/10.1103/PhysRevD.83.014018}{\doi{10.1103/PhysRevD.83.014018}},
  \href{http://www.arXiv.org/abs/1010.1869}{\texttt{arXiv:1010.1869}}.

\bibitem{Barone:2002cv}
V.~Barone and E.~Predazzi, ``High-energy particle diffraction'', volume 565 of
  \textit{ Texts and Monographs in Physics}.
\newblock Springer-Verlag, Berlin Heidelberg, 2002.
\newblock ISBN~978-3-540-42107-8.

\bibitem{ATLASFRG}
\hrefCMSnoop {}{{ATLAS Collaboration}, ``Rapidity gap cross sections measured
  with the {ATLAS} detector in $pp$ collisions at {$\sqrt{s}=$} 7 {TeV}'',}
  \textit{ Eur. Phys. J. C} \textbf{ 72} (2012) 1926,
  \href{http://dx.doi.org/10.1140/epjc/s10052-012-1926-0}{\doi{10.1140/epjc/s10052-012-1926-0}},
  \href{http://www.arXiv.org/abs/1201.2808}{\texttt{arXiv:1201.2808}}.

\bibitem{Robert}
\hrefCMSnoop {}{{CMS Collaboration}, ``Measurement of diffraction dissociation
  cross sections in pp collisions at {$\sqrt{s}=$} 7 {TeV}'',} \textit{ Phys.
  Rev. D} \textbf{ 92} (2015) 012003,
  \href{http://dx.doi.org/10.1103/PhysRevD.92.012003}{\doi{10.1103/PhysRevD.92.012003}},
  \href{http://www.arXiv.org/abs/1503.08689}{\texttt{arXiv:1503.08689}}.

\bibitem{Baltz:2007kq}
\hrefCMSnoop {}{A.~J. Baltz {et~al.}, ``The physics of ultraperipheral
  collisions at the {LHC}'',} \textit{ Phys. Rept.} \textbf{ 458} (2008) 1,
  \href{http://dx.doi.org/10.1016/j.physrep.2007.12.001}{\doi{10.1016/j.physrep.2007.12.001}},
  \href{http://www.arXiv.org/abs/0706.3356}{\texttt{arXiv:0706.3356}}.

\bibitem{Luna:2004eg}
\hrefCMSnoop {}{R.~Luna, A.~Zepeda, C.~A. Garcia~Canal, and S.~J. Sciutto,
  ``Influence of diffractive interactions on cosmic ray air showers'',}
  \textit{ Phys. Rev. D} \textbf{ 70} (2004) 114034,
  \href{http://dx.doi.org/10.1103/PhysRevD.70.114034}{\doi{10.1103/PhysRevD.70.114034}},
  \href{http://www.arXiv.org/abs/hep-ph/0408303}{\texttt{arXiv:hep-ph/0408303}}.

\bibitem{Ostapchenko:2014mna}
\hrefCMSnoop {}{S.~Ostapchenko, ``{LHC} data on inelastic diffraction and
  uncertainties in the predictions for longitudinal extensive air shower
  development'',} \textit{ Phys. Rev. D} \textbf{ 89} (2014) 074009,
  \href{http://dx.doi.org/10.1103/PhysRevD.89.074009}{\doi{10.1103/PhysRevD.89.074009}},
  \href{http://www.arXiv.org/abs/1402.5084}{\texttt{arXiv:1402.5084}}.

\bibitem{Gribov:1967vfb}
\hrefCMSnoop {}{V.~N. Gribov, ``A reggeon diagram technique'',} \textit{ Sov.
  Phys. JETP} \textbf{ 53} (1967) 654.

\bibitem{Sjostrand:1987su}
\hrefCMSnoop {}{T.~Sj{\"o}strand and M.~van Zijl, ``A multiple interaction
  model for the event structure in hadron collisions'',} \textit{ Phys. Rev. D}
  \textbf{ 36} (1987) 2019,
  \href{http://dx.doi.org/10.1103/PhysRevD.36.2019}{\doi{10.1103/PhysRevD.36.2019}}.

\bibitem{HEPData}
\hrefCMSnoop {}{}{HEPData} record for this analysis, 2022.
\newblock
  \href{http://dx.doi.org/10.17182/hepdata.88293}{\doi{10.17182/hepdata.88293}}.

\bibitem{Sirunyan:2020zal}
\hrefCMSnoop {}{{CMS Collaboration}, ``Performance of the {CMS} {Level-1}
  trigger in proton-proton collisions at {$\sqrt{s}=$} 13 {TeV}'',} \textit{
  JINST} \textbf{ 15} (2020) P10017,
  \href{http://dx.doi.org/10.1088/1748-0221/15/10/P10017}{\doi{10.1088/1748-0221/15/10/P10017}},
  \href{http://www.arXiv.org/abs/2006.10165}{\texttt{arXiv:2006.10165}}.

\bibitem{Khachatryan:2016bia}
\hrefCMSnoop {}{{CMS Collaboration}, ``The {CMS} trigger system'',} \textit{
  JINST} \textbf{ 12} (2017) P01020,
  \href{http://dx.doi.org/10.1088/1748-0221/12/01/P01020}{\doi{10.1088/1748-0221/12/01/P01020}},
  \href{http://www.arXiv.org/abs/1609.02366}{\texttt{arXiv:1609.02366}}.

\bibitem{Chatrchyan:2008zzk}
\hrefCMSnoop {}{{CMS Collaboration}, ``The {CMS} experiment at the {CERN}
  {LHC}'',} \textit{ JINST} \textbf{ 3} (2008) S08004,
  \href{http://dx.doi.org/10.1088/1748-0221/3/08/S08004}{\doi{10.1088/1748-0221/3/08/S08004}}.

\bibitem{Sirunyan:2017ulk}
\hrefCMSnoop {}{{CMS Collaboration}, ``Particle-flow reconstruction and global
  event description with the {CMS} detector'',} \textit{ JINST} \textbf{ 12}
  (2017) P10003,
  \href{http://dx.doi.org/10.1088/1748-0221/12/10/P10003}{\doi{10.1088/1748-0221/12/10/P10003}},
  \href{http://www.arXiv.org/abs/1706.04965}{\texttt{arXiv:1706.04965}}.

\bibitem{Agostinelli:2002hh}
\hrefCMSnoop {}{{{\GEANTfour}} Collaboration, ``{\GEANTfour}---a simulation
  toolkit'',} \textit{ Nucl. Instrum. Meth. A} \textbf{ 506} (2003) 250,
  \href{http://dx.doi.org/10.1016/S0168-9002(03)01368-8}{\doi{10.1016/S0168-9002(03)01368-8}}.

\bibitem{CMS:2018fkg}
\href {https://cds.cern.ch/record/2628652}{{CMS Collaboration}, ``{CMS}
  luminosity measurement using 2016 proton-nucleus collisions at
  nucleon-nucleon center-of-mass energy of 8.16 {TeV}'',} CMS Physics Analysis
  Summary CMS-PAS-LUM-17-002, 2018.

\bibitem{CMS_TR}
\hrefCMSnoop {}{{CMS Collaboration}, ``Description and performance of track and
  primary-vertex reconstruction with the {CMS} tracker'',} \textit{ JINST}
  \textbf{ 9} (2014) P10009,
  \href{http://dx.doi.org/10.1088/1748-0221/9/10/P10009}{\doi{10.1088/1748-0221/9/10/P10009}},
  \href{http://www.arXiv.org/abs/1405.6569}{\texttt{arXiv:1405.6569}}.

\bibitem{Khachatryan:2016qhq}
\hrefCMSnoop {}{{CMS Collaboration}, ``Coherent $\mathrm{J}/\psi$
  photoproduction in ultra-peripheral {PbPb} collisions at
  {$\sqrt{\smash[b]{s_{_{\mathrm{NN}}}}}=$} 2.76 {TeV} with the {CMS}
  experiment'',} \textit{ Phys. Lett. B} \textbf{ 772} (2017) 489,
  \href{http://dx.doi.org/10.1016/j.physletb.2017.07.001}{\doi{10.1016/j.physletb.2017.07.001}},
  \href{http://www.arXiv.org/abs/1605.06966}{\texttt{arXiv:1605.06966}}.

\bibitem{Sirunyan:2019nog}
\hrefCMSnoop {}{{CMS Collaboration}, ``Measurement of exclusive $\rho(770)^0$
  photoproduction in ultraperipheral {pPb} collisions at
  {$\sqrt{\smash[b]{s_{_{\mathrm{NN}}}}}=$} 5.02 {TeV}'',} \textit{ Eur. Phys.
  J. C} \textbf{ 79} (2019) 702,
  \href{http://dx.doi.org/10.1140/epjc/s10052-019-7202-9}{\doi{10.1140/epjc/s10052-019-7202-9}},
  \href{http://www.arXiv.org/abs/1902.01339}{\texttt{arXiv:1902.01339}}.

\bibitem{Sirunyan:2018sav}
\hrefCMSnoop {}{{CMS Collaboration}, ``Measurement of exclusive {\PgU}
  photoproduction from protons in {pPb} collisions at
  {$\sqrt{\smash[b]{s_{_{\mathrm{NN}}}}}=$} 5.02 {TeV}'',} \textit{ Eur. Phys.
  J. C} \textbf{ 79} (2019) 277,
  \href{http://dx.doi.org/10.1140/epjc/s10052-019-6774-8}{\doi{10.1140/epjc/s10052-019-6774-8}},
  \href{http://www.arXiv.org/abs/1809.11080}{\texttt{arXiv:1809.11080}}.

\bibitem{Sirunyan:2018fhl}
\hrefCMSnoop {}{{CMS Collaboration}, ``Evidence for light-by-light scattering
  and searches for axion-like particles in ultraperipheral {PbPb} collisions at
  {$\sqrt{\smash[b]{s_{_{\mathrm{NN}}}}}=$} 5.02 {TeV}'',} \textit{ Phys. Lett.
  B} \textbf{ 797} (2019) 134826,
  \href{http://dx.doi.org/10.1016/j.physletb.2019.134826}{\doi{10.1016/j.physletb.2019.134826}},
  \href{http://www.arXiv.org/abs/1810.04602}{\texttt{arXiv:1810.04602}}.

\bibitem{Khachatryan:2016odn}
\hrefCMSnoop {}{{CMS Collaboration}, ``Charged-particle nuclear modification
  factors in {PbPb} and {pPb} collisions at
  {$\sqrt{\smash[b]{s_{_{\mathrm{NN}}}}}=$} 5.02 {TeV}'',} \textit{ JHEP}
  \textbf{ 04} (2017) 039,
  \href{http://dx.doi.org/10.1007/JHEP04(2017)039}{\doi{10.1007/JHEP04(2017)039}},
  \href{http://www.arXiv.org/abs/1611.01664}{\texttt{arXiv:1611.01664}}.

\bibitem{Sirunyan:2019djp}
\hrefCMSnoop {}{{CMS Collaboration}, ``Calibration of the {CMS} hadron
  calorimeters using proton-proton collision data at {$\sqrt{s}=$} 13 {TeV}'',}
  \textit{ JINST} \textbf{ 15} (2020) P05002,
  \href{http://dx.doi.org/10.1088/1748-0221/15/05/P05002}{\doi{10.1088/1748-0221/15/05/P05002}},
  \href{http://www.arXiv.org/abs/1910.00079}{\texttt{arXiv:1910.00079}}.

\bibitem{DAgostini:1994fjx}
\hrefCMSnoop {}{G.~D'Agostini, ``A multidimensional unfolding method based on
  {Bayes'} theorem'',} \textit{ Nucl. Instrum. Meth. A} \textbf{ 362} (1995)
  487,
  \href{http://dx.doi.org/10.1016/0168-9002(95)00274-X}{\doi{10.1016/0168-9002(95)00274-X}}.

\bibitem{Adye:2011gm}
\hrefCMSnoop {}{T.~Adye, ``Unfolding algorithms and tests using {RooUnfold}'',}
  in \textit{ Proceedings, {PHYSTAT} 2011 Workshop on Statistical Issues
  Related to Discovery Claims in Search Experiments and Unfolding}, p.~313.
\newblock CERN, Geneva, Switzerland, 2011.
\newblock \href{http://www.arXiv.org/abs/1105.1160}{\texttt{arXiv:1105.1160}}.
\newblock
  \href{http://dx.doi.org/10.5170/CERN-2011-006.313}{\doi{10.5170/CERN-2011-006.313}}.

\bibitem{Schmitt:UnfoldingMethods}
\hrefCMSnoop {}{S.~Schmitt, ``Data unfolding methods in high energy physics'',}
  in \textit{ EPJ Web Conf. XIIth Quark Confinement and the Hadron Spectrum},
  volume 137, p.~11.
\newblock 2017.
\newblock
  \href{http://www.arXiv.org/abs/1611.01927}{\texttt{arXiv:1611.01927}}.
\newblock
  \href{http://dx.doi.org/10.1051/epjconf/201713711008}{\doi{10.1051/epjconf/201713711008}}.

\bibitem{Suranyi:2019mgm}
\hrefCMSnoop {}{O.~Sur{\'a}nyi, ``Performance of the {CMS} {Zero Degree
  Calorimeters} in the 2016 {pPb} run'',} \textit{ J. Phys. Conf. Ser.}
  \textbf{ 1162} (2019) 012005,
  \href{http://dx.doi.org/10.1088/1742-6596/1162/1/012005}{\doi{10.1088/1742-6596/1162/1/012005}}.

\bibitem{Suranyi:2021ssd}
\hrefCMSnoop {}{O.~Sur{\'a}nyi {et~al.}, ``Performance of the {CMS} {Zero
  Degree Calorimeters} in {pPb} collisions at the {LHC}'',} \textit{ JINST}
  \textbf{ 16} (2021) P05008,
  \href{http://dx.doi.org/10.1088/1748-0221/16/05/P05008}{\doi{10.1088/1748-0221/16/05/P05008}},
  \href{http://www.arXiv.org/abs/2102.06640}{\texttt{arXiv:2102.06640}}.

\end{thebibliography}\endgroup
\cleardoublepage \appendix\section{The CMS Collaboration \label{app:collab}}\begin{sloppypar}\hyphenpenalty=5000\widowpenalty=500\clubpenalty=5000
\cmsinstitute{Yerevan Physics Institute, Yerevan, Armenia}
{\tolerance=6000
A.~Tumasyan\cmsAuthorMark{1}\cmsorcid{0009-0000-0684-6742}
\par}
\cmsinstitute{Institut f\"{u}r Hochenergiephysik, Vienna, Austria}
{\tolerance=6000
W.~Adam\cmsorcid{0000-0001-9099-4341}, F.~Ambrogi\cmsorcid{0000-0002-9486-0444}, T.~Bergauer\cmsorcid{0000-0002-5786-0293}, M.~Dragicevic\cmsorcid{0000-0003-1967-6783}, J.~Er\"{o}, A.~Escalante~Del~Valle\cmsorcid{0000-0002-9702-6359}, R.~Fr\"{u}hwirth\cmsAuthorMark{2}\cmsorcid{0000-0002-0054-3369}, M.~Jeitler\cmsAuthorMark{2}\cmsorcid{0000-0002-5141-9560}, N.~Krammer\cmsorcid{0000-0002-0548-0985}, L.~Lechner\cmsorcid{0000-0002-3065-1141}, D.~Liko\cmsorcid{0000-0002-3380-473X}, T.~Madlener\cmsorcid{0000-0002-0128-6536}, I.~Mikulec\cmsorcid{0000-0003-0385-2746}, F.M.~Pitters, N.~Rad, J.~Schieck\cmsAuthorMark{2}\cmsorcid{0000-0002-1058-8093}, R.~Sch\"{o}fbeck\cmsorcid{0000-0002-2332-8784}, M.~Spanring\cmsorcid{0000-0001-6328-7887}, S.~Templ\cmsorcid{0000-0003-3137-5692}, W.~Waltenberger\cmsorcid{0000-0002-6215-7228}, C.-E.~Wulz\cmsAuthorMark{2}\cmsorcid{0000-0001-9226-5812}, M.~Zarucki\cmsorcid{0000-0003-1510-5772}
% ...
\par}
\vskip\cmsinstskip
\dag:~Deceased\\
$^{1}$Also at Yerevan State University, Yerevan, Armenia\\
$^{2}$Also at TU Wien, Vienna, Austria\\
% ...
\end{sloppypar}
\end{document}